\documentclass[conference]{IEEEtran}
\IEEEoverridecommandlockouts
\usepackage{cite}
\usepackage{amsmath,amssymb,amsfonts}
\usepackage{algorithmic}
\usepackage{graphicx}
\usepackage{multirow}
\usepackage{caption}
\usepackage{subcaption}
\usepackage{textcomp}
\usepackage{xcolor}
\usepackage{balance}

\usepackage[pdftex]{hyperref}
\usepackage{tikz}
\newcommand\copyrighttext{%
    \footnotesize \textcopyright 2024 IEEE. Personal use of this material is permitted.
    Permission from IEEE must be obtained for all other uses, in any current or future
    media, including reprinting/republishing this material for advertising or promotional
    purposes, creating new collective works, for resale or redistribution to servers or
    lists, or reuse of any copyrighted component of this work in other works.
    DOI: \href{https://doi.org/10.1109/CLUSTER59578.2024.00039}{https://doi.org/10.1109/CLUSTER59578.2024.00039}}
\newcommand\copyrightnotice{%
    \begin{tikzpicture}[remember picture,overlay]
        \node[anchor=south,yshift=10pt] at (current page.south) {\fbox{\parbox{\dimexpr\textwidth-\fboxsep-\fboxrule\relax}{\copyrighttext}}};
    \end{tikzpicture}%
}

\def\BibTeX{{\rm B\kern-.05em{\sc i\kern-.025em b}\kern-.08em
    T\kern-.1667em\lower.7ex\hbox{E}\kern-.125emX}}
\begin{document}

\newcommand{\dockeywords}{Resource Management, Scientific Workflow, Memory Allocation, Memory Prediction, Machine Learning}

\title{Sizey: Memory-Efficient Execution of  \\  Scientific Workflow Tasks}

\author{
    \IEEEauthorblockN{Jonathan Bader\IEEEauthorrefmark{1}, Fabian Skalski\IEEEauthorrefmark{1}, Fabian Lehmann\IEEEauthorrefmark{2},
        Dominik Scheinert\IEEEauthorrefmark{1}, \\ Jonathan Will\IEEEauthorrefmark{1}, Lauritz Thamsen\IEEEauthorrefmark{3}, and Odej Kao\IEEEauthorrefmark{1}}

    \IEEEauthorblockA{
        \IEEEauthorrefmark{1}
        \{firstname.lastname\}@tu-berlin.de, Technische Universität Berlin, Germany\\
    }
    \IEEEauthorblockA{
        \IEEEauthorrefmark{2}
        fabian.lehmann@informatik.hu-berlin.de, Humboldt-Universität zu Berlin, Germany\\
    }
    \IEEEauthorblockA{
        \IEEEauthorrefmark{3}
        lauritz.thamsen@glasgow.ac.uk, University of Glasgow, United Kingdom\\
    }

}

\maketitle
\copyrightnotice
\begin{abstract}
As the amount of available data continues to grow in fields as diverse as bioinformatics, physics, and remote sensing, the importance of scientific workflows in the design and implementation of reproducible data analysis pipelines increases.
When developing workflows, resource requirements must be defined for each type of task in the workflow.
Typically, task types vary widely in their computational demands because they are simply wrappers for arbitrary black-box analysis tools.
Furthermore, the resource consumption for the same task type can vary considerably as well due to different inputs.
Since underestimating memory resources leads to bottlenecks and task failures, workflow developers tend to overestimate memory resources.
However, overprovisioning of memory wastes resources and limits cluster throughput.

Addressing this problem, we propose \emph{Sizey}, a novel online memory prediction method for workflow tasks.
During workflow execution, Sizey simultaneously trains multiple machine learning models and then dynamically selects the best model for each workflow task.
To evaluate the quality of the model, we introduce a novel resource allocation quality (RAQ) score based on memory prediction accuracy and efficiency.
Sizey's prediction models are retrained and re-evaluated online during workflow execution, continuously incorporating metrics from completed tasks.

Our evaluation with a prototype implementation of Sizey uses metrics from six real-world scientific workflows from the popular nf-core framework and shows a median reduction in memory waste over time of 24.68\% compared to the respective best-performing state-of-the-art baseline.

\end{abstract}

\IEEEpeerreviewmaketitle


\section{Introduction}\label{sec:INTRO}
Scientific workflow management systems (SWMS) such as Nextflow~\cite{nextflow}, Pegasus~\cite{pegasus}, or Snakemake~\cite{koster2012snakemake} enable the creation of reproducible data analysis workflows.
SWMS help scientists from various domains, such as genomics, remote sensing, and material science~\cite{nf_core,lehmannFORCENextflowScalable2021,schaarschmidt2021workflow,liew2016scientific,coleman2022wfcommons,macaw}, to abstract execution-specific details and to cope with huge amounts of data.
Workflows are often defined as a directed acyclic graph (DAG), consisting of a set of black-box task types $B$ and a set of directed edges $E$.
Each task type $b$ serves as a wrapper for a black-box analysis tool and as a template for physical task instances $T$ with concrete inputs.
Each individual edge $e$ connects two tasks and expresses their dependency, dataflow, and execution order.

SWMS submit physical task instances to a cluster resource manager such as Kubernetes~\cite{kubernetes}, Slurm~\cite{slurm}, or HTCondor~\cite{condor}.
To assign each physical task to a suitable cluster node, resource managers rely on adequate resource requirements, such as the number of processor cores and main memory required.
Typically, task type resource requirements are provided by users and workflow developers and are essentially estimates~\cite{ilyushkin2018impact,feitelson2015workload,hirales2012multiple}. 
However, estimating the resource requirements of tasks is known to be difficult and error prone~\cite{baderPredictDynamicMemoryRequ2023,tovar2022dynamic,witt2019learning,witt2019predictive}.
At the same time, accurate estimates of especially the required memory are crucial. 
The underestimation of memory can result in bottlenecks, costly offloading to disks, and even task failures.
Therefore, users are incentivized to overestimate resource usage to avoid partial or complete restarts of workflows.
However, requesting more memory than needed wastes precious cluster resources and effectively limits parallelism within and across workflow applications.

\begin{figure}[t]
	\centering
	\includegraphics[width=\columnwidth]{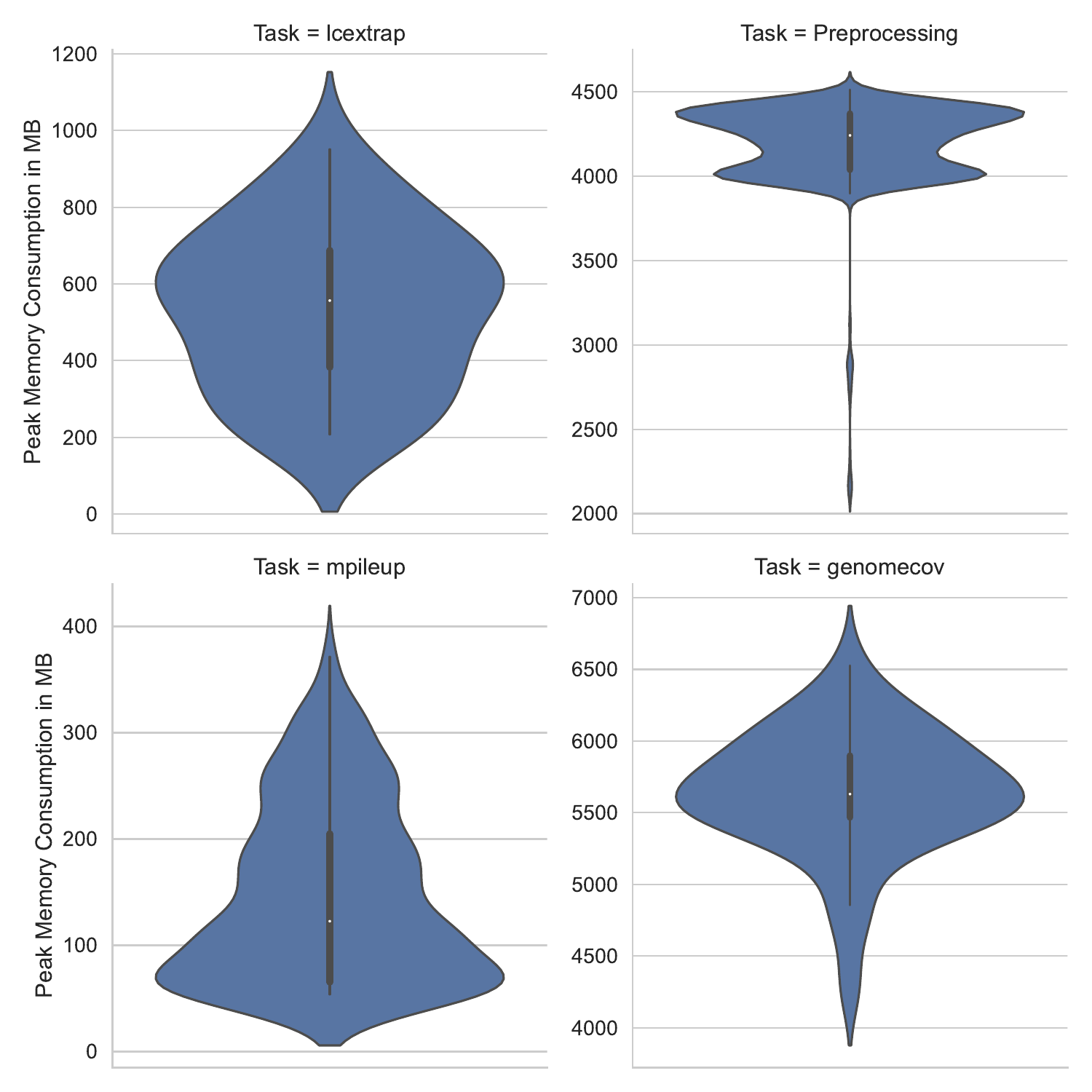}
	\caption{Distribution of peak memory consumption of four task types,  each executed repeatedly with varying input sizes.}
	\label{fig:var_mem_con}
\end{figure}

The memory consumption of executing different physical tasks that use the same task template but different inputs can vary substantially and often depends on the input data.
Figure~\ref{fig:var_mem_con} shows the distribution of memory consumption for four different task types that were executed repeatedly with varying input sizes.
It can be observed that the memory consumption for executions of the same task type differs significantly. Furthermore, the memory consumption also varies between the different task types. 
For example, the \emph{lcextrap} task instances consume between 200MB and 1GB with a median of approximately 550MB.
Consequently, the prediction of the peak memory usage of task instances should be automated, taking into account the differences between task types, the variances in memory consumption between their concrete physical task executions, as well as the impact of input data.
Further, incorporating resource measurements online, i.e., during the execution of the workflow, allows refining prediction models during runtime.

\begin{figure}[t]
	\centering
	\includegraphics[width=\columnwidth]{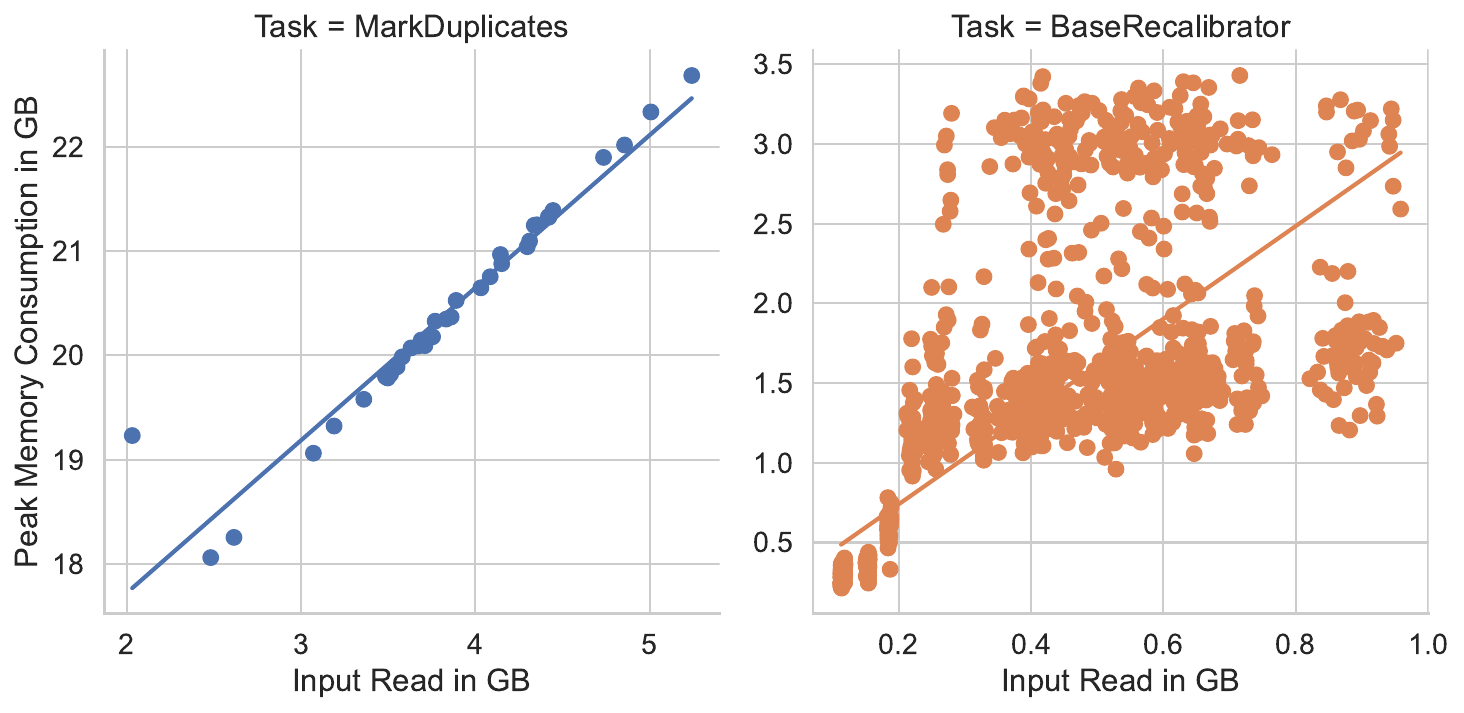}
	\caption{Memory consumption in relation to the input read of the physical task instances of two different task types, with a linear regression applied to these data points.}
	\label{fig:non_linear}
\end{figure}

The majority of existing methods for predicting memory requirements in scientific workflows employ machine learning to address the challenge of memory prediction~\cite{witt2019predictive}.
However, these methods neglect that not each workflow task can be modeled well with the same type of machine learning model.
For instance, while many workflow tasks yield a linear memory consumption behavior correlating with the input data size, this is not the case for all.
Figure~\ref{fig:non_linear} shows how the memory consumption of tasks depends on input reads.
It can be seen that there is a clear linear correlation for \emph{MarkDuplicates}.
However, using a linear model for the \emph{BaseRecalibrator} task would lead to half of the task instances failing due to insufficient memory, and the other half would waste significant memory due to overestimates.
Therefore, an adaptive prediction method is needed that selects an adequate machine learning model for all task types.

This paper presents Sizey, a novel online task memory prediction method for scientific workflows that employs the best out of multiple machine learning models for each task. 
Upon submission of a task, Sizey searches a provenance database for previous task executions of the same task type. 
If the task is of an unknown task type, it is submitted directly to the resource manager, resorting to the user-provided, usually conservative memory estimate. 
On completion of the task, the monitoring data are stored in the provenance database.
Based on the previous executions, Sizey trains a set of diverse machine learning models in parallel, with each model's prediction yielding a memory estimate.
Subsequently, the prediction performance of the models is assessed using a new resource allocation quality (RAQ) score, which measures both the observed accuracy and the estimated resource efficiency.
Based on the RAQ score, Sizey selects the best-performing prediction model for each task and adjusts the prediction with a dynamic fault tolerance offset to minimize costly task failures caused by underestimates.
Finally, as new data become available after the completion of each task, a lightweight - and thus fast - online learning step updates the existing models.

\noindent\textit{Contributions.} This paper makes the following contributions:

\begin{itemize}
    \item We present our novel task memory prediction method, Sizey, which employs a set of diverse machine learning models to adaptively predict a task's memory in order to reduce memory resource wastage.
    \item We provide an open-source implementation of our method Sizey, as an easily extendable interface\footnote{github.com/dos-group/sizey}.
    \item We evaluate Sizey using six large-scale workflows from the nf-core repository~\cite{nf_core} and compare it to four state-of-the-art methods. The experimental results demonstrate that Sizey outperforms all state-of-the-art baselines with a median memory wastage reduction of at least 24.68\% compared to all baselines.
\end{itemize}
\iffalse
\emph{Outline}. The remainder of the paper is structured as follows.
Section~\ref{sec:RELATED_WORK} discusses the related work.
Section~\ref{sec:APPROACH} describes our online memory prediction method Sizey.
Section~\ref{sec:Eval1} provides an explanation of the experimental setup employed in this study, including the scientific workflows, baselines, and cluster setup utilized. It also presents the results of a comprehensive evaluation conducted with four state-of-the-art baselines, gives insights into Sizey's prediction model, and offers a discussion of these results.
Section~\ref{sec:CONCLUSION} summarizes this paper and provides a short outlook toward future work.
\fi

\section{Approach}\label{sec:APPROACH}
In this section, we first provide an overview of our proposed method.
Second, we explain the granularity of the prediction model used by Sizey and the selection of the model classes. 
Third, we explain our resource allocation quality (RAQ) score that combines model accuracy and resource efficiency.
Fourth, we describe how we weight and combine the prediction results into a final prediction.
Lastly, we explain our initial sizing strategy and handling of task failures due to underprovisioning.

\begin{figure*}[t!]
	\centering
	\includegraphics[width=1.0\textwidth]{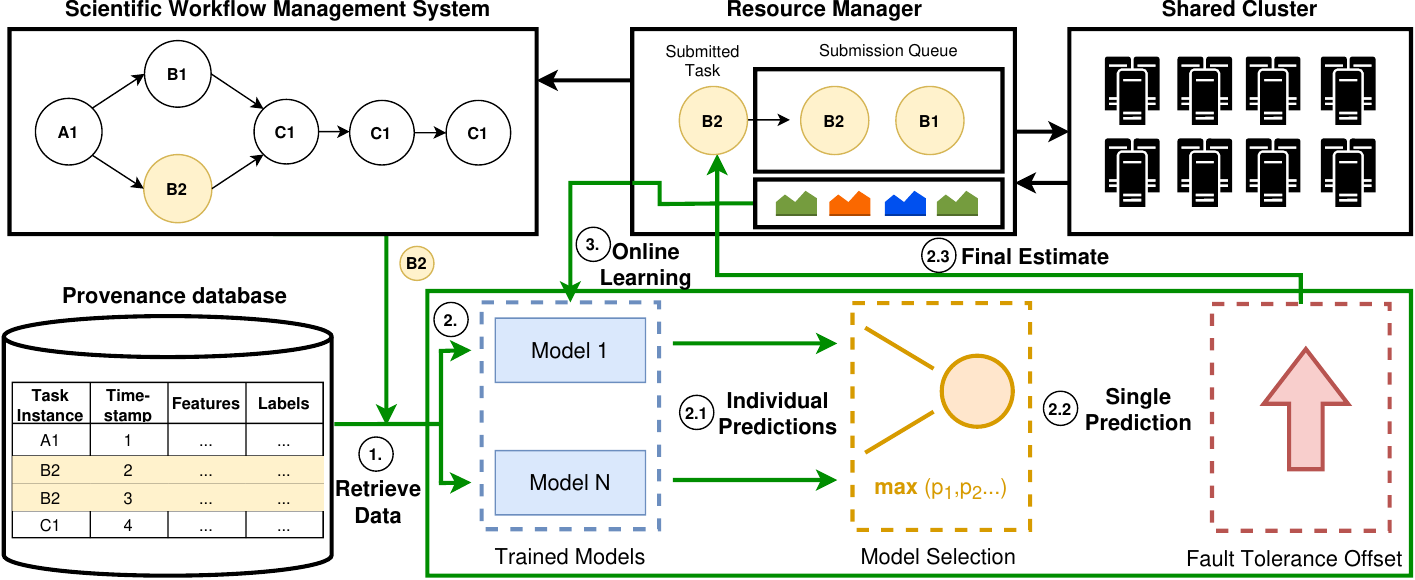}
	\caption{The figure provides an overview of how Sizey (green) is used in a scientific workflow execution environment. Sizey uses data from a provenance database and trains multiple machine learning models in parallel. The individual predictions are then combined into a single prediction that also includes an offset to ensure sufficient memory is allocated for tasks executing on the cluster. During workflow execution and after task completion, the model is retrained and reevaluated.}
	\label{fig:overview}
\end{figure*}

\subsection{Overview}
Sizey estimates the peak memory consumption of a submitted task instance in an online manner during the execution of a workflow. 
Figure \ref{fig:overview} illustrates how this works.

\paragraph{Data Retrieval}
In Phase \textcircled{\raisebox{-0.9pt}{1}}, a task is submitted for execution, which causes Sizey to retrieve provenance details from the provenance database connected to the SWMS, such as the number of running tasks and information about historical task executions, including the task name and resource consumption.

\paragraph{Memory Prediction}
In Phase~\textcircled{\raisebox{-0.9pt}{2}}, Sizey estimates the memory consumption of the task to be executed.
For this, in Step~\textit{2.1}, a set of diverse machine learning models is used, where each predictor produces a resource estimate.
Next, our resource allocation quality (RAQ) score compares the accuracy and efficiency of the individual predictors.
In Step~\textit{2.2}, a gating mechanism selects the models and, depending on the strategy, decides between aggregating and maximizing the individual predictions according to their RAQ scores to produce a single prediction.
In Step \textit{2.3}, this single prediction is adjusted based on the selected offset strategy and how often the task execution was attempted before.
Subsequently, the task instance and resource prediction are submitted to the cluster resource manager for executing the task with the determined memory.

\paragraph{Online Learning}
In Phase~\textcircled{\raisebox{-0.9pt}{3}}, a task has been completed and new training data becomes available.
Therefore, in this online learning phase, all models are updated. 
The update steps are efficient due to the lightweight inputs and models used.

\paragraph{Assumptions}
We make the following assumptions for predicting the memory consumption of workflow tasks:

\begin{itemize}
    \item [A1:] We assume that workflows are used on many input files, thus involving many data-parallel task executions and repeated executions of the same task type, allowing online learning during workflow execution.
    \item [A2:] We assume that scheduling, which involves ordering task instances and assigning them to cluster nodes, is the responsibility of the resource manager and is therefore outside the scope of this work. 
    \item [A3:] We assume that the resource manager enforces strict resource limits on memory allocations, resulting in a failed task execution when exceeding these limits.
\end{itemize}

\subsection{Model Granularity, Classes, and Features}
\begin{figure}[t!]
	\centering
	\includegraphics[width=\columnwidth]{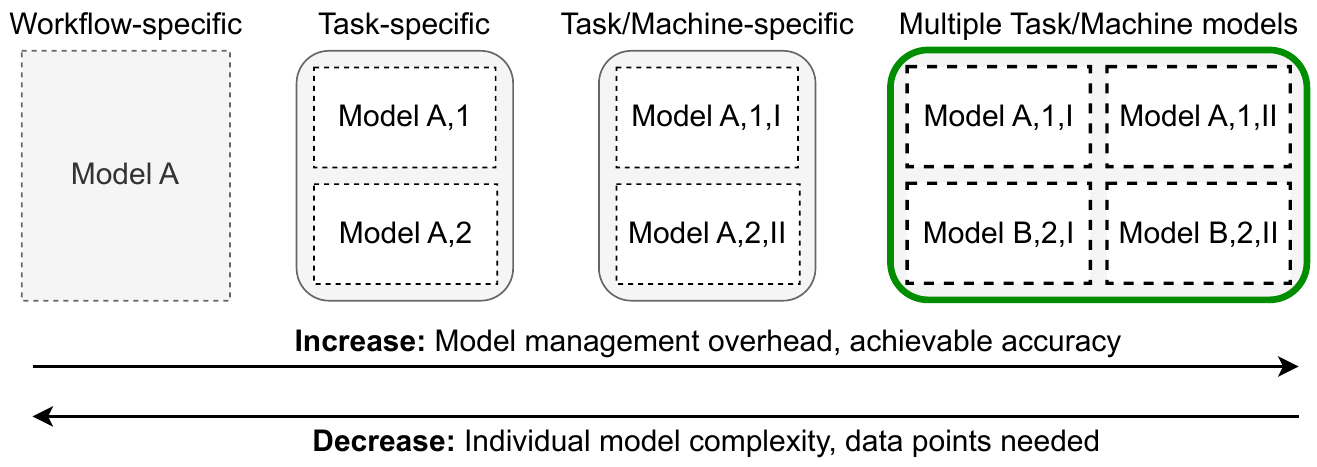}
	\caption{ Granularity levels for a given workflow with model \textbf{A}, task types \textbf{1, 2} on machines \textbf{I} and \textbf{II}; Sizey method in bold green box.}
	\label{fig:model_granularity}
\end{figure}

To achieve the goal of memory wastage reduction, precise models for resource predictions are needed.
%Figure~\ref{fig:model_granularity} explains the relation between model granularity and prediction accuracy.
%Generally, fine-granular models achieve better accuracy compared to coarse-grained counterparts since less noise caused by differt machine configurations is present~\cite{singh2017machine}.
%However, they require more and fine-grained training data and include a more sophisticated model management.
Sizey uses multiple fine-granular models to predict the memory consumption.  
This technique is visualized in Figure~\ref{fig:model_granularity}, differentiating between a workflow-specific model, a task type specific model, a task/machine-specific model, and our proposed model granularity.
Sizey employs the most fine-granular model by incorporating several models for each task-machine configuration.
The rationale behind this approach is that tasks exhibit heterogeneous computational patterns that vary even more with different machine configurations. 
This implies that a model type that performs well on one task type may perform poorly on another.

\begin{figure}[t!]
	\centering
	\includegraphics[width=\columnwidth]{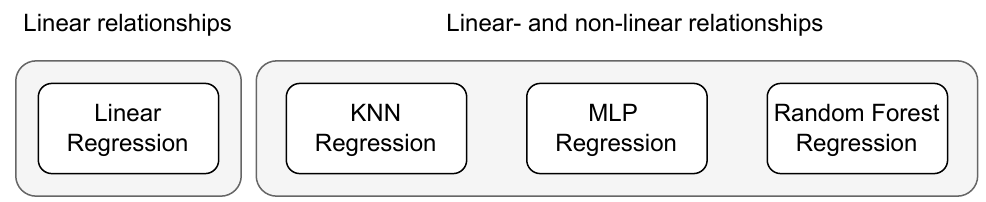}
	\caption{The four model classes used by Sizey for modeling linear and non-linear relationships between input data sizes and peak memory consumption.}
	\label{fig:model_classes}
\end{figure}

Figure~\ref{fig:model_classes} shows the model classes incorporated by Sizey. 
As stated in~\cite{bux2017hi,witt2019learning, bader2024lotaru}, tasks frequently yield a linear relationship between input data size and memory consumption.
We exploit this observation by integrating a linear regression model class into the pool of used model classes. 
However, as presented in the introduction, the relationship between input data size and memory consumption does not have to be linear.
The k-nearest neighbors (KNN) regression model allows historical observations similar to the task currently estimated to influence the resource prediction.
We include MLP regression to accurately model more complex, nonlinear relationships, such as memory usage that grows as the square of the amount of input data.
Incorporating a random forest regression model makes our method more resistant to overfitting, especially when there are not many historical task executions.

%Table~\ref{tab:features} describes the input features Sizey uses for and for the  prediction model inputs.
%The runtime features include task metadata, dynamic input data, and task failure tracking.
%For example, with the arrival timestamp some models can identify temporal task dependencies, while the retry counter identifies a failed sizing.
%The environment features include machine information and cluster environment features.
%The number of running tasks can give insights to the current cluster usage and the machine information defines a single machine in the cluster.
%We characterize a machine by the number of available CPU cores and amount of main memory
%The historical labels include monitoring information from previous executions regarding CPU and memory usage.

Since Sizey runs online, there is no classical segmentation into training and test data during workflow execution. 
Instead, the prediction accuracy of individual models is permanently assessed.
When new task resource measurements become available as tasks finish, each model performs an update step. 
Such an update step can also be a complete retraining of the models if 1) the training time is short, 2) the time until the next task submission is long enough for a complete retraining, or 3) a complete retraining for better performance is desired.

\subsection{Resource Allocation Quality (RAQ)}
Sizey makes use of a set of machine learning models.
This poses the question of how to select the best-fitting model or even a model combination for certain abstract task-machine configurations.

The RAQ score describes the presumed goodness of the task's resource estimate provided by one associated trained model. 
The composite score is derived from two subscores: the accuracy score and the efficiency score. All three scores, the RAQ score, the accuracy score, and the efficiency score, are normalized scalars between 0 (worst) and 1 (best).

\paragraph{Accuracy Score}
The accuracy score expresses a model's average prediction performance.
Let $T=\{t^{(i)}|i\in \mathbb{N}_0,\text{ and } 0\leq i \leq S-1\}$ be the set of historical task instances, with $S$ denoting the size of the set.
The resource estimate for a historical task instance $t^{(j)}$ given by the $i$-th model is denoted as $\hat{y}_{i,t^{(j)}}$, while $y_{t^{(j)}}$ denotes the actual resource consumption of $t^{(j)}$. 
Then, the accuracy score $AS_{i,t^*}$ of the $i$-th model associated with the combination of task type and machine type for the current task $t^*$ can be calculated as follows:

\begin{equation} 
	\label{equ:accuracy_score}
	AS_{i,t^*} = \frac{1}{S} \sum_{\forall t^{(j)}\in T} \Big( 1 - min(|\frac{\hat{y}_{i,t^{(j)}} - y_{t^{(j)}}}{y_{t^{(j)}}}|, 1) \Big).
\end{equation}

In Equation~\ref{equ:accuracy_score}, the individual error terms are bounded at one to prohibit large estimation outliers from skewing the computed scores.
The resulting accuracy score describes the mean prediction performance of a model across the set of task instances.
A mean prediction error of zero indicates a perfectly accurate prediction and would result in an accuracy score of 1, the best possible value. 
Accuracy scores are updated over time, while models predict and learn from new task data.

\paragraph{Efficiency Score}
The efficiency score compares a model's individual memory predictions to the other model's memory predictions. 
With $N$ being the number of models available for the desired combination of a task type and a machine configuration and $\hat{y}_{i,t^*}$ the prediction of the $i$-th model for the current task $t^*$, its dynamic efficiency score $ES_{i,t^*}$ is calculated as follows:

\begin{equation} 
	\label{equ:efficiency_score}
	ES_{i,t^*} = 1 - \frac{\hat{y}_{i,t^*}}{max_{j=1,...,N}(\hat{y}_{j,t^*})}.
\end{equation}

Efficiency scores are model output-specific and punish outlying resource estimates that are huge in magnitude relative to other model outputs. 
These large estimation outliers may occur in various cases, for instance, when extrapolation occurs during the early training stages. 
By validating individual estimate magnitudes with those from other predictors, the risk of wasting anomalously large amounts of memory resources is mitigated.
The efficiency score associated with the largest resource estimate is always 0. 

\paragraph{Computing the Resource Allocation Quality}
Our RAQ score includes one mandatory hyperparameter $\alpha$.
With $\alpha \in [0,1]$ and $AS_{i,t^*}$ and $ES_{i,t^*}$ being the respective accuracy and efficiency scores, the resource allocation quality $RAQ_{i,t^*}$ of the $i$-th model for the current task $t^*$ can be calculated as: 

\begin{equation} 
	\label{equ:raq_formula}
	RAQ_{i,t^*} = (1 - \alpha) \cdot AS_{i,t^*} + \alpha \cdot ES_{i,t^*}.
\end{equation}

When $\alpha$ is close to zero, accurate models are favored, whereas values close to one punish large outlying estimates more strongly.

\subsection{Model Selection Strategy}

\begin{figure}[]
\includegraphics[width=1.0\columnwidth]{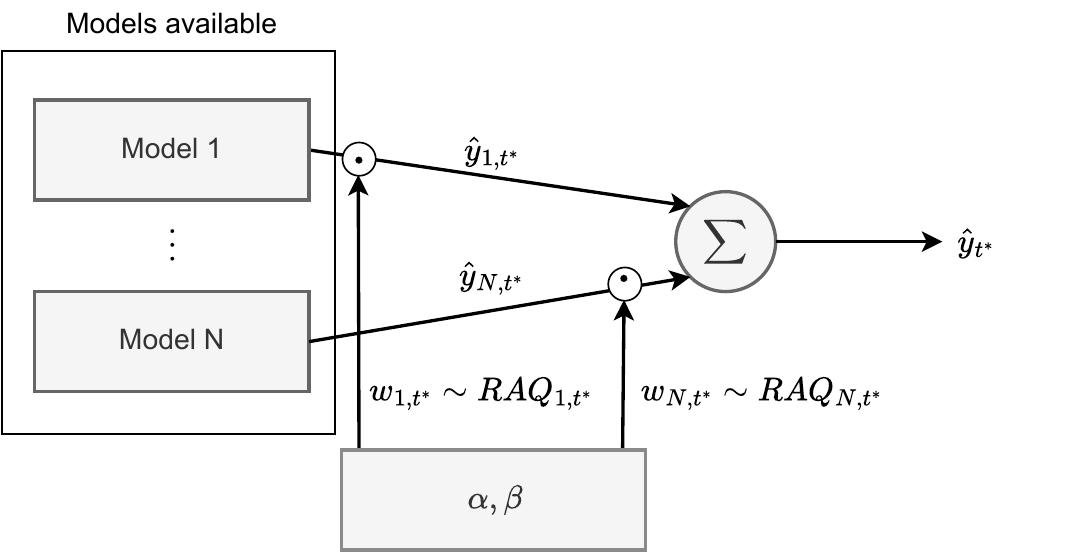}
	\caption{Role of resource allocation quality scores in model selection and weighting}
	\label{fig:gating_mechanism}
\end{figure}

Figure~\ref{fig:gating_mechanism} shows the structure of our proposed gating mechanism that is used to select the models that contribute to the final prediction.
Similar to a gating network from the mixture of experts domain~\cite{masoudnia2014mixture,shazeer2017outrageously}, our gating mechanism assigns individual weights to the outputs of the individual predictors. 
These weights depend on our RAQ score that we introduced before. 

Sizey can be used with two different gating strategies.
The gating strategy determines the predictor's weights from Figure~\ref{fig:gating_mechanism}. 

\paragraph{Argmax Strategy}

The first strategy is called \textit{Argmax} and weights all predictors with zero except the one corresponding to the highest RAQ score. 
In that case, the output of the associated predictor is chosen as the aggregate resource estimate $\hat{y}_{t^*}$. 

\paragraph{Interpolation Strategy}
Whenever the \textit{interpolation} strategy is enabled, the hyperparameter $\beta \in [1, \infty)$ becomes mandatory and weights are calculated using softmax: 

\begin{equation} 
	\label{equ:gating_predictor_interpolation}
	\hat{y}_{t^*} = \sum_{i=1}^{N} \Big( \hat{y}_{i,t^*} \cdot w_{i,t^*} \Big) = \sum_{i=1}^{N} \Big( \hat{y}_{i,t^*} \cdot \frac{e^{\beta \cdot RAQ_{i,t^*}}}{\sum_{j=1}^{N} e^{\beta \cdot RAQ_{j,t^*}}} \Big).
\end{equation}

While the Argmax strategy is more opportunistic towards the presumably best available model, the interpolation strategy seeks to utilize a model output consensus.

\subsection{Prediction Offset and Failure Handling}

Since Sizey aims to predict memory as accurately as possible, small underpredictions can quickly lead to task failures.
Therefore, we use a dynamic offset strategy that increases Sizey's aggregate memory prediction for the task.
As with the models, our dynamic offset strategy chooses among several possible offsets: the standard deviation, the standard deviation of underpredictions, the median prediction error, and the median prediction error of underpredictions.
During online learning, Sizey selects the offset that would have caused the least wastage based on the tasks already executed.

If a task instance still fails due to underprediction, the maximum amount of task memory ever observed is allocated.
For each subsequent attempt to size a previously failed task instance, the given resource estimate is continuously doubled until the machine resources are exhausted.

\section{Evaluation}\label{sec:Eval1}
Our evaluation section is divided into five subsections: experimental setup, baselines, experimental results, Sizey model insights, and the discussion of the results. 

\subsection{Experimental Setup}

\begin{figure*}[t]
	\centering
	\includegraphics[width=1.0\textwidth]{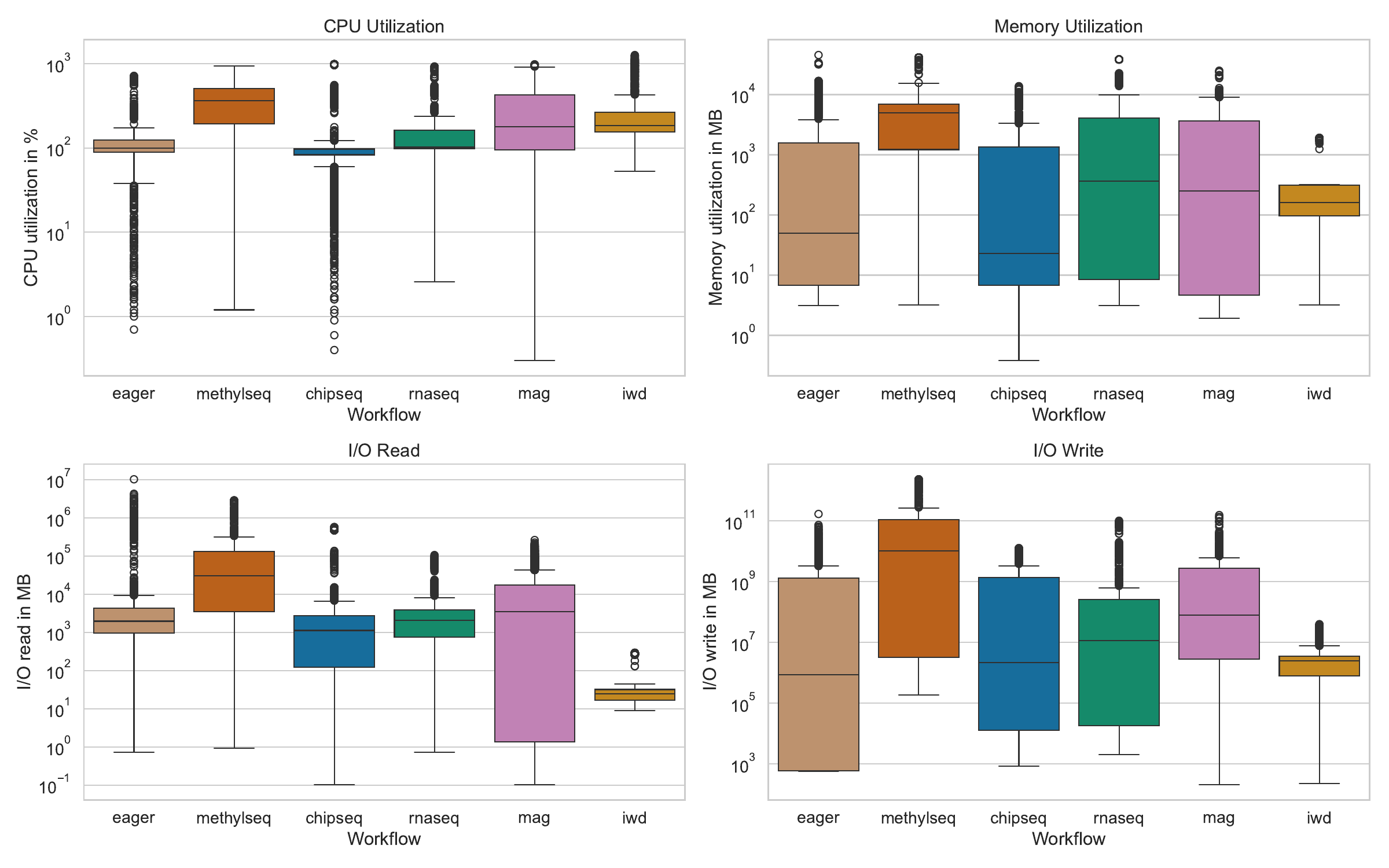}
	\caption{Distribution of memory, CPU, and I/O utilization of the executed workflows on a logarithmic y-axis scale.}
	\label{fig:workflow_profiles}
\end{figure*}

For our experiments, we executed and measured six real-world workflows on our cluster infrastructure, five workflows from the bioinformatics domain, and one workflow from the field of remote sensing:

\begin{enumerate}
    \item For the \textbf{eager workflow}~\cite{yates2021reproducible}, we use input data published in a study in 2018 by de Barros Damgaard et al. on the population history of the Eurasian steppe~\cite{damgaard2018137}.  
    \item For the \textbf{rnaseq workflow}\footnote{github.com/nf-core/rnaseq}, we use data from a bladder cancer cells study~\cite{green2021molecular}.
    \item For the \textbf{mag workflow}\footnote{github.com/nf-core/mag}, we use input data from a study of the gut microbiome~\cite{karlsson2013gut}.
    \item For the \textbf{chipseq workflow}\footnote{github.com/nf-core/chipseq} we use data from a prostate cancer study ~\cite{baumgart2020darolutamide}.
    \item The input data for the \textbf{methylseq workflow}\footnote{github.com/nf-core/methylseq} are from a study of human pluripotent stem cells in 2020 by Thompson et al.~\cite{thompson2020low}.
    \item The \textbf{iwd workflow}~\cite{rettelbach2022images} uses more than 300 images within and around fire scars in Western Alaska and uses computer vision and graph analysis to examine the landscapes.
\end{enumerate}

For data collection, we run these workflows on an eight-node Kubernetes cluster, each equipped with an AMD EPYC 7282 16-core 32-thread processor, 128GB of DDR4 memory,
and two 960GB SATA III SSDs.
The Kubernetes cluster uses a Ceph file system for data storage.

Figure \ref{fig:workflow_profiles} illustrates the memory, CPU, and I/O resource consumption of the experimental workflows.
One can see that all workflows yield different resource usage patterns.
For instance, the methylseq workflow yields many I/O-intensive tasks, while also being CPU-intensive.
Table~\ref{tab:workflow_instances} illustrates the number of task types and the average number of task instances per task type on a workflow basis. 
The rnaseq and the chipseq workflow have the greatest number of task types, yet the lowest average number of task instances per task type. 
In contrast, the iwd workflow has the lowest number of task types, yet the second highest average number of task instances per task type.
Using such a heterogeneous set of workflows and task types allows us to test the heterogeneous computing landscape of shared cluster environments.

To enable a comprehensive evaluation of memory prediction of large-scale workflow executions with multiple methods, we simulate an online environment where our measured real-world metrics from completed task executions can be incorporated into the learning process, similar to as in a real-world system.
Simulation parameters can be used to define aspects such as the amount of historical data, failure strategies, model parameters, etc. 
An important parameter we will use in our experiments is the time-to-failure value, which expresses at what point in time a task will fail.
For example, a time-to-failure value of 1.0 means that we expect tasks to fail at the end of their execution.
We filter out very short-running tasks and tasks with a single or few executions.

Our implementation is written in Python 3.11 and uses scikit-learn, numpy, and scipy as the main libraries and keeps the models in memory.
For our simulation experiments, we used all four machine learning models: linear regression, knn regression, MLP regression, and random forest regression.
We have implemented a version that always fully retrains the models and a version that provides incremental updates and caches the best hyperparameters over the workflow execution.
In addition, we select an $\alpha$ value of 0.0 and use the Interpolation strategy for Sizey.
A comparison of the full model training and the incremental updates, as well as an analysis of selecting the parameters, can be found in the model insights evaluation (Section~\ref{subsec:sizey_model}).

\begin{table}
\centering
\caption{The table shows the average number of task instances per task type for each experimental workflow.}
\label{tab:workflow_instances}
\begin{tabular}{l|r|r}

Workflow & \# Task Types & AVG \# Task Instances per Task Type \\ \hline
eager &    13       & 121 \\ \hline
methylseq &    9       & 100 \\ \hline
chipseq &     30      & 82 \\ \hline
rnaseq &  30    &     39 \\ \hline
mag &    8      & 720 \\ \hline
iwd &    5      & 332 \\ \hline

\end{tabular}
\end{table}

\subsection{Baselines}

\begin{figure*}[t!]
\centering
\begin{subfigure}[t]{1\columnwidth}
   \centering
  \includegraphics[width=1\columnwidth]{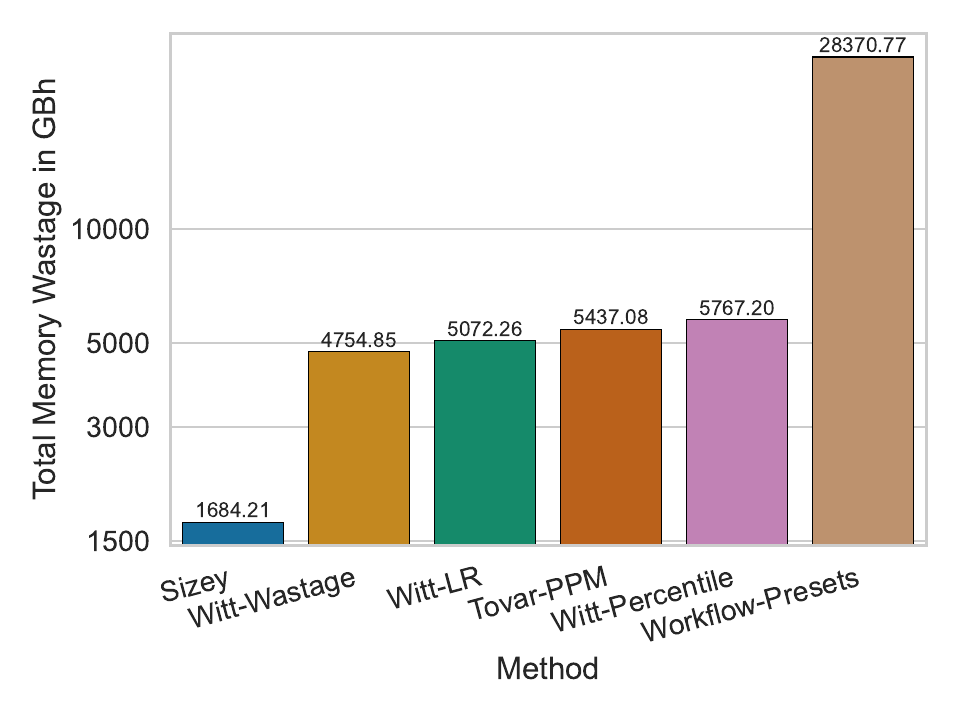}
  \caption{Wastage over time in gigabyte-hours (GBh) aggregated over all workflows with a time-to-failure of 1.0 on a logarithmic y-axis scale.}
  \label{fig:eval_wastage_ttf1}
\end{subfigure}
\begin{subfigure}[t]{1\columnwidth}
   \centering
  \includegraphics[width=1\columnwidth]{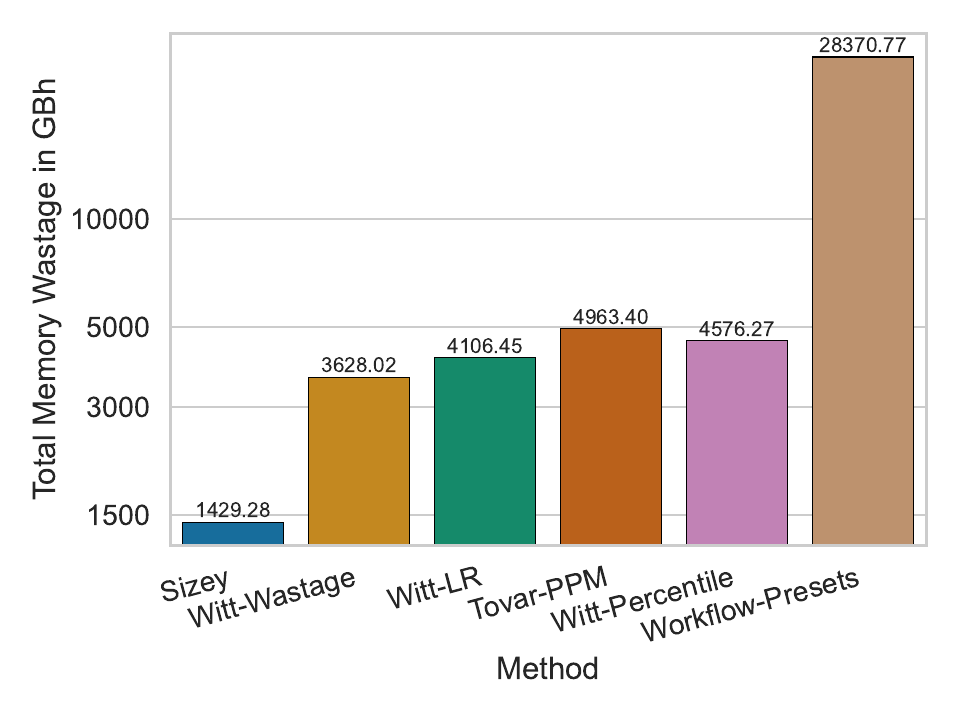}
  \caption{Wastage over time in gigabyte-hours (GBh) aggregated over all workflows with a time to failure of 0.5 on a logarithmic y-axis scale.}
  \label{fig:eval_wastage_ttf05}
\end{subfigure}
\begin{subfigure}[t]{1\columnwidth}
   \centering
  \includegraphics[width=1\columnwidth]{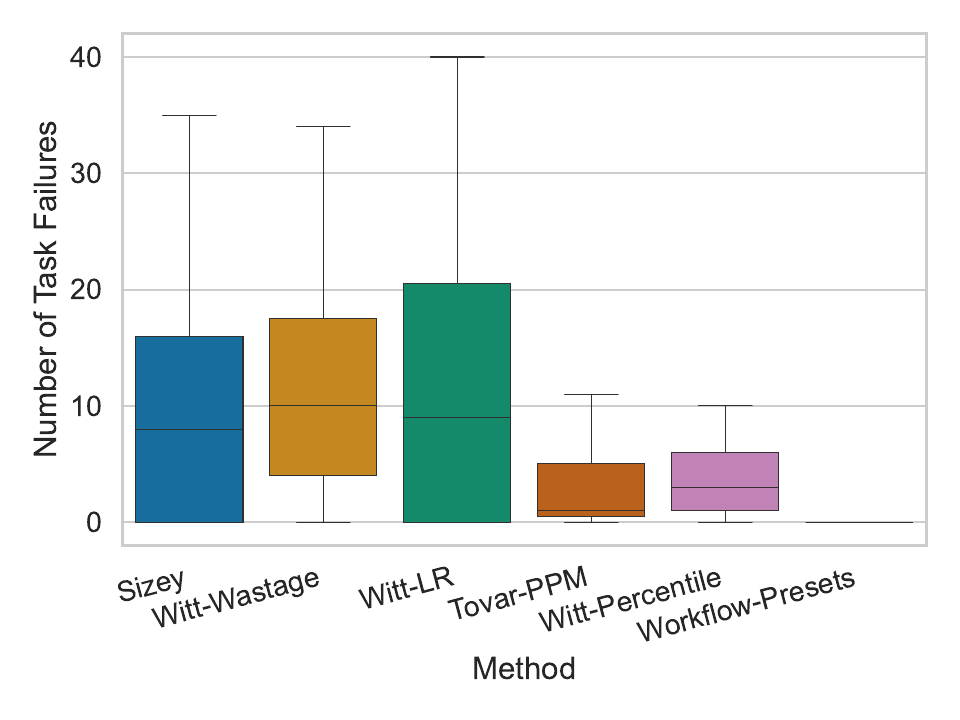}
  \caption{Distribution of the methods' task failures, aggregated by task type, with outliers hidden.}
  \label{fig:eval_failures}
\end{subfigure}
\begin{subfigure}[t]{1\columnwidth}
   \centering
  \includegraphics[width=1\columnwidth]{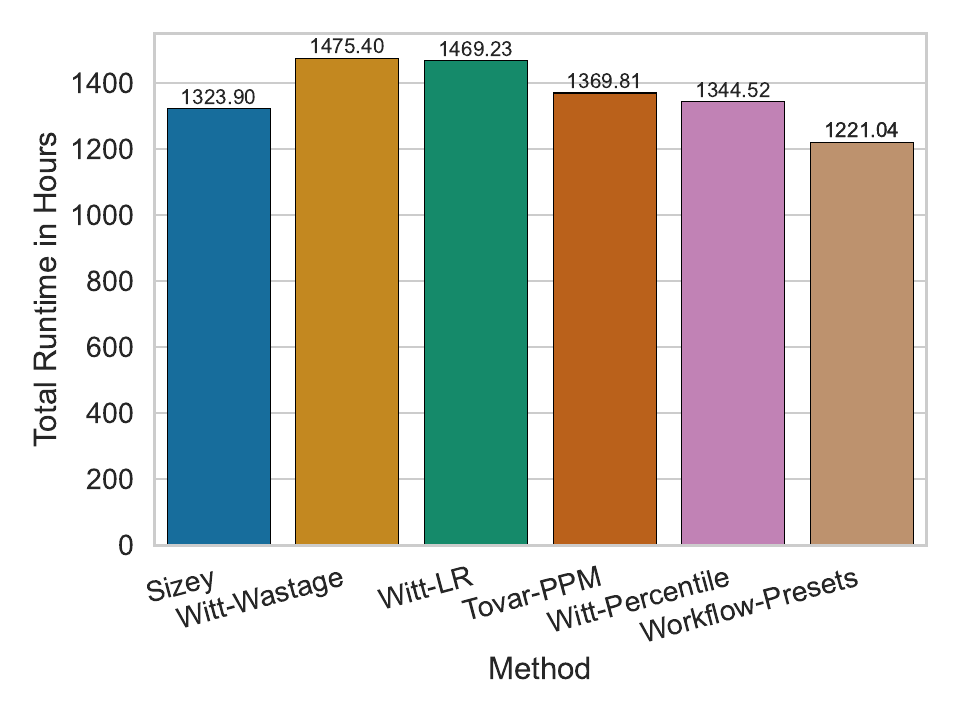}
  \caption{Aggregated workflow task runtimes for each method.}
  \label{fig:eval_time}
\end{subfigure}
\caption{Experimental results of predicting memory consumption for tasks of six workflows.}
\label{fig:eval_total}
\end{figure*}%

Our method is evaluated against four state-of-the-art baselines along with the default workflow setups provided by the workflow developers, which serve as a sanity baseline.

Tovar et al.~\cite{tovar2017job} use historical peak memory probabilities (\textbf{Tovar-PPM}) to determine the initial memory allocation for workflow tasks. 
Their strategy aims to reduce the overall probability that resource peaks will exceed the allocated memory. 
Should the initial allocation underestimate the required resource, resulting in task failure, Tovar et al. allocate a node's maximum memory. 
We used the authors' source code and integrated it into our setup.

Witt et al.~\cite{witt2019learning} propose a low-wastage regression that optimizes the resource wastage instead of the prediction error and is also applied in an online fashion.
Their method is based on a linear model and doubles the predicted memory upon task failure.
We use the source code provided on the author's Github account and refer to the method as \textbf{Witt-Wastage}.

Witt et al.~\cite{witt2019feedback} use two different methods using an online learning mode.
The percentile predictor predicts the percentile peak memory usage of all historical tasks. 
The authors propose a conservative estimate, using the 95th percentile to avoid task failures.
As a second method, they propose a linear regression (LR), using the input size as a feature and adding an offset on the prediction.
We refer to these methods as \textbf{Witt-Percentile} and \textbf{Witt-LR}.
Given the absence of accessible source code, our implementation of the method was based on the descriptions provided in the paper.

\subsection{Experimental Results}

\begin{table*}[t]
\centering
\caption{Aggregated memory wastage over time for all workflows and methods evaluated.}
\label{tab:wastetime}
\begin{tabular}{l|r|r|r|r|r|r}

Method & methylseq & chipseq & eager & rnaseq & mag & iwd \\ \hline
Sizey & \textbf{631.62} & \textbf{79.38} & 678.19 & \textbf{43.62} & \textbf{251.05} & \textbf{0.36} \\
Witt-Wastage & 3565.11 & 214.60 & \textbf{491.16} & 176.39 & 323.62 & 0.55 \\
Witt-LR & 988.90 & 136.33 & 3585.19 & 57.91 & 301.00 & 2.94 \\
Tovar-PPM & 4080.60 & 211.02 & 624.14 & 195.26 & 309.36 & 16.70 \\
Witt-Percentile & 4372.19 & 94.70 & 860.16 & 128.90 & 309.81 & 1.44 \\
Workflow-Presets & 22596.14 & 260.61 & 2304.53 & 1238.62 & 1955.01 & 15.86 \\

\end{tabular}
\end{table*}

Figure~\ref{fig:eval_total} shows the aggregated results of the memory prediction for six different workflows, including the memory wastage over time for two different time-to-failure values, the distribution of task failures, and the aggregated task runtimes.

Figure~\ref{fig:eval_wastage_ttf1} shows the aggregated wastage over time with a time-to-failure value of 1.0 for all methods in gigabyte-hours. 
Among all methods, the Workflow-Presets show the highest wastage, approximately 17 times higher than Sizey and 6 times higher than the best-performing baseline.
Sizey exhibits the lowest wastage over time for all methods and achieves a 64.58\% lower wastage compared to the best-performing baseline, Witt-Wastage.
Also, all other baselines are able to outperform the Workflow-Presets but still show significantly higher wastage over time compared to Sizey.

Figure~\ref{fig:eval_wastage_ttf05} also shows the aggregated wastage over time for all methods in gigabyte-hours.
However, we now configure a time-to-failure value of 0.5, which means that tasks will fail earlier during execution.
Here we can see that Sizey also shows the lowest wastage, a decrease of 60.60\% compared to the Witt-Wastage baseline.
In contrast to Figure~\ref{fig:eval_wastage_ttf1}, the relative difference in wastage between Sizey and the Witt-Wastage method has slightly decreased.
Again, the Workflow-Presets show the highest wastage and also the same wastage as before as they are not impacted by the time-to-failure value.
It is also observed that all state-of-the-art methods benefit from a lower time-to-failure and that Witt-Percentile outperforms Tovar-PPM with tasks failing faster.

Figure~\ref{fig:eval_failures} depicts the distribution of task failures aggregated by task type. 
Witt-Wastage exhibits the highest median number of task failures followed by Witt-LR and Sizey.
The methods' whiskers and the difference between the first and the third quartile indicate a high variability in the number of failures. 
Witt-Percentile and Tovar-PPM exhibit a comparable low number of failures, which is anticipated given that the Witt-Percentile method employs a highly conservative 95th percentile predictor and Tovar-PPM uses a very conservative failure handling.
Because workflow defaults are user-defined estimates set to prevent task failures, task failures do not occur.

Figure~\ref{fig:eval_time} exhibits the aggregated task runtimes without considering the memory wastage.
Here, the Workflow-Presets achieve the lowest aggregate.
This is expected as we did not observe any task failures, resulting in no mandatory task restarts.
Sizey achieves the second-lowest aggregated task runtime, slightly followed by Witt-Percentile.
The Witt-Wastage method shows the highest runtime, which can be explained by the high number of average task failures.

Table~\ref{tab:wastetime} shows the wastage on a workflow basis for all methods and experimental workflows.
We can observe that in five out of six workflows, Sizey achieves the lowest wastage over time and a considerable improvement to the best-performing competitor.
The highest relative difference compared to the best-performing baseline can be observed for the methylseq workflow, showing a reduction of memory wastage of 36.13\%, followed by the iwd workflow with a reduction of 34.55\%.
Notably, the performance of the baselines fluctuates, and no baseline is consistently second best after Sizey.
Four out of six times, the Workflow-Presets result in the highest wastage over time; for the eager workflow, the Workflow-Presets outperform Witt-LR, and for the iwd workflow they outperform Tovar-PPM.

\subsection{Sizey Model Insights}
\label{subsec:sizey_model}

\begin{figure}[t!]
	\centering
	\includegraphics[width=\columnwidth]{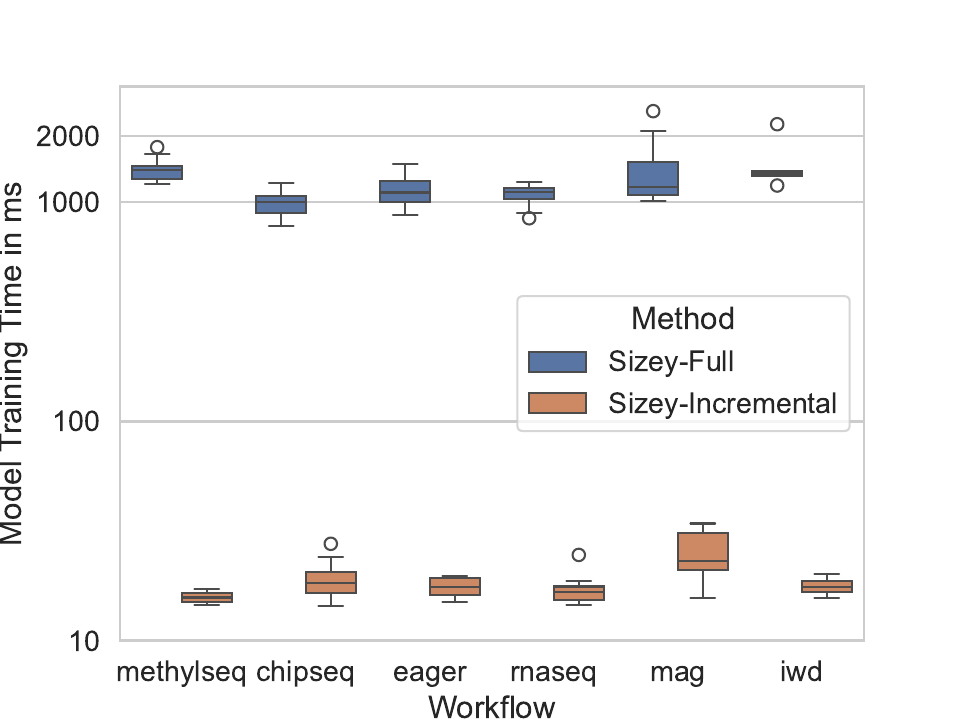}
	\caption{Time required to train Sizey for full retraining and incremental retraining, including model hyperparameter optimization. y-axis uses a logarithmic scale.}
	\label{fig:training-time}
\end{figure}

\begin{figure}[t!]
	\centering
	\includegraphics[width=\columnwidth]{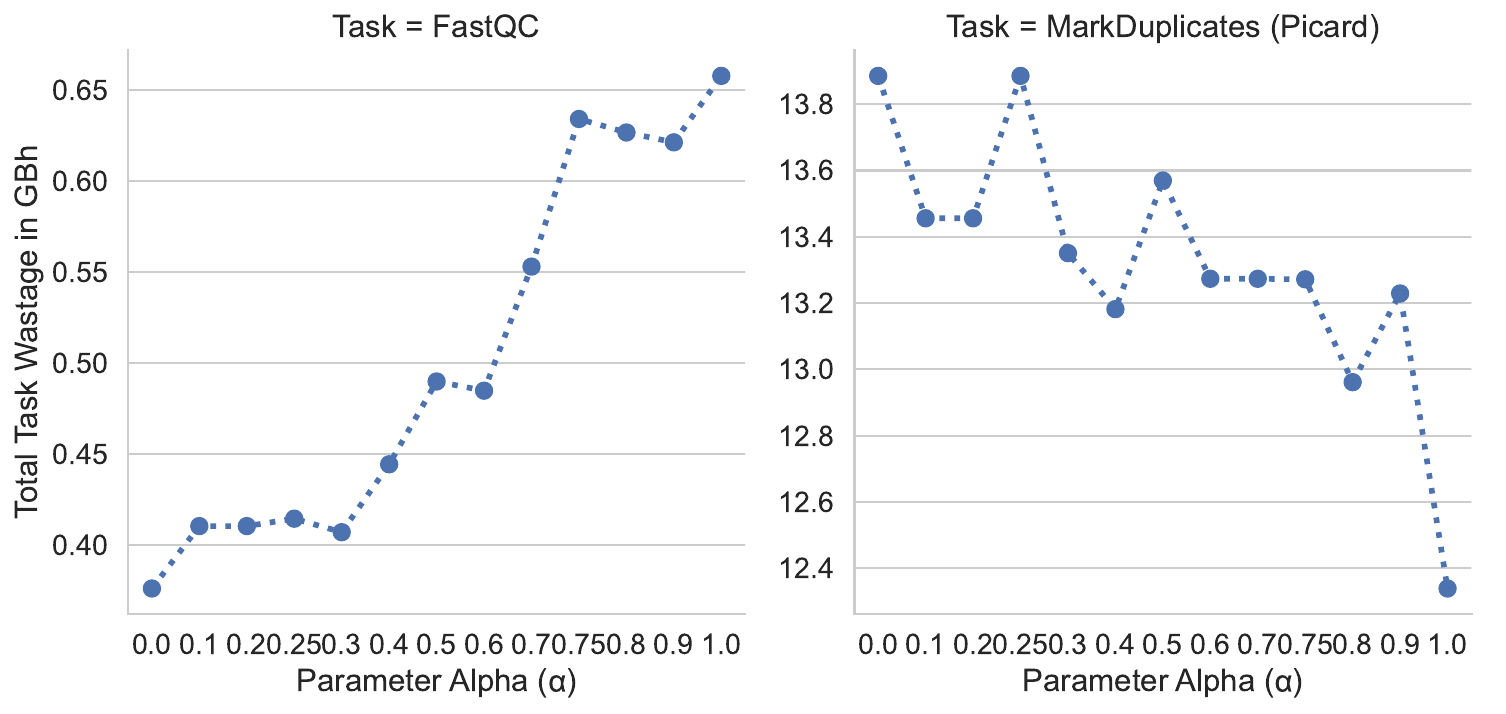}
	\caption{Impact of the parameter alpha on wastage over time in GBh for two rnaseq tasks.}
	\label{fig:alpha}
\end{figure}

\begin{figure}[t!]
	\centering
	\includegraphics[width=\columnwidth]{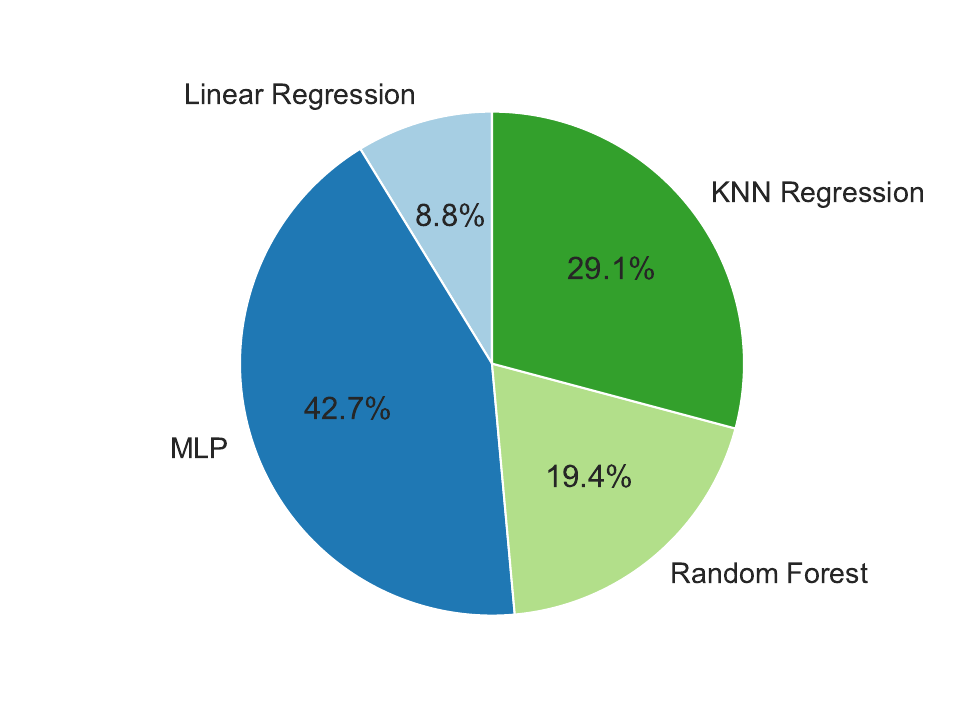}
	\caption{Proportion of model classes selected by Sizey for the rnaseq workflow.}
	\label{fig:model_distro}
\end{figure}

Doing a complete retraining using Sizey to predict a task's memory consumption, including hyperparameter optimization, we observed a median training time of 1.09 seconds.
Using an incremental approach, we achieved a reduction of 98.39\% with a median training time of 17.5 milliseconds.
Figure~\ref{fig:training-time} illustrates these findings and demonstrates that the training time for full retraining is comparable across all workflows. 
A similar behavior can be observed for the incremental training.
Further, we observed that using Sizey with incremental training increased its median wastage over time by 6.1\%.

Figure~\ref{fig:alpha} exemplifies the impact of alpha when Sizey is executed with an alpha value ranging from 0 to 1.0 for two rnaseq workflow tasks.
We can observe that the FastQC task shows a tendency for wastage to be lower for a lower alpha parameter.
In contrast, the MarkDuplicates task shows the opposite pattern with decreasing wastage as alpha increases.
We believe this observation is due to the definition of our efficiency score, which favors models with lower predictions and penalizes possible overprediction outliers to achieve more accurate predictions.
However, this may also lead to more task failures as it favors lower predictions.
Also, in general, the additional data we have analyzed do not show a clear tendency to choose a higher or lower alpha.

Figure~\ref{fig:model_distro} shows the distribution of selected model classes across all experimental workflows when Sizey is run with the Argmax strategy.
It can be observed that the MLP regressor is employed for 42.7\% of the task memory predictions, followed by KNN with 29.1\%, Random Forest with 19.4\%, and Linear Regression with 8.8\%.
The linear model is a good fit when only a few training data points are available, for instance, at the beginning of a workflow execution.
Once more data become available, Sizey can switch to more complex models.

\begin{figure}[t!]
	\centering
	\includegraphics[width=\columnwidth]{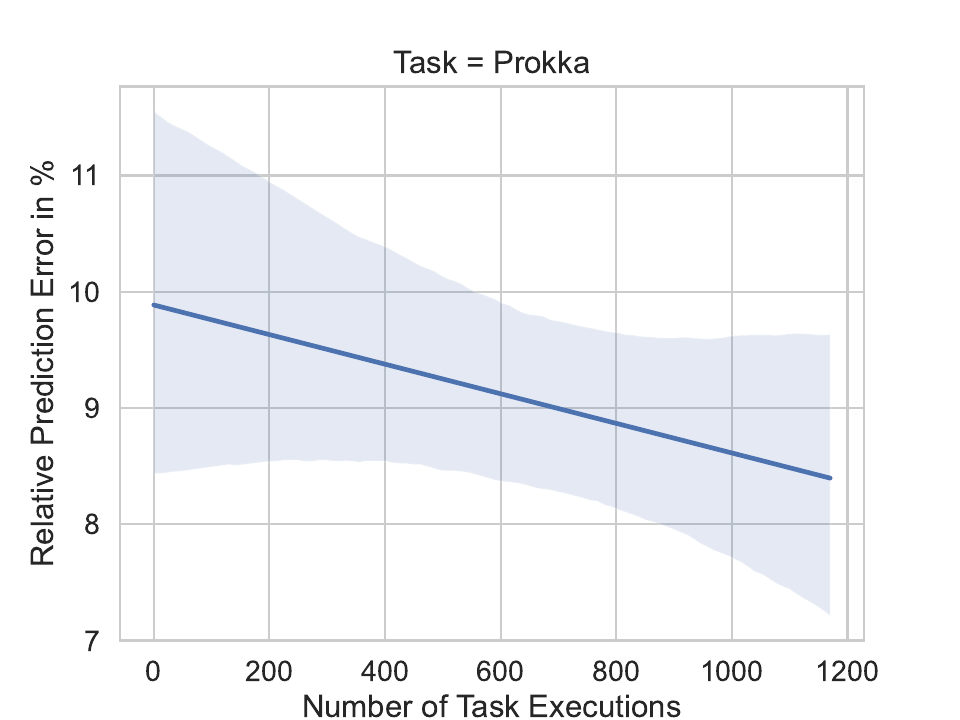}
	\caption{Trend of Sizey's relative memory prediction error over time (without offsetting) when running 1171 Prokka task instances.}
	\label{fig:Prokka_pred_error}
\end{figure}

In Figure~\ref{fig:Prokka_pred_error}, we show as an example Sizey's relative memory prediction error for the Prokka task from the mag workflow.
This is the raw prediction error, without any offsetting, so simply the difference between the predicted memory and the actual memory usage.
The blue line shows the regression of the data points with a 95 percent confidence interval.
We can observe, both for the regression trend line and for the confidence interval, that the prediction error decreases with the number of task executions due to the online learning.
We observed this behavior for the majority of task executions with Sizey, but not with all baselines.

\subsection{Discussion}

In our experimental results, we observed that Sizey is able to significantly reduce memory wastage compared to all state-of-the-art baselines.
While Sizey shows the largest reduction in memory wastage for the methylseq workflow at 36.13\% compared to the best-performing baseline, the overall reduction across all workflows is about 60.60\% compared to the best-performing baseline.
This is because the performance of the baselines varies greatly for each workflow, while Sizey achieves low memory wastage for all workflows.

Our experimental results show that a reduction in memory wastage is accompanied by an increase in task execution time, due to the occurrence of failed tasks. 
Although the additional time is relatively small, especially when compared to the memory savings, it is a factor to be considered. 
One approach to reducing task failures and therefore time, while still reducing memory, is to choose a more conservative offset.

Our experiments show that Sizey's lightweight models allow a quick training time.
Using Sizey with an incremental model reduces the training time further while still significantly outperforming the baselines.
The analysis of the alpha parameter showed no clear trend.
For some workflow tasks, a lower alpha results in less wastage, while for others it will be the other way around.
We assume that this is due to the definition of our efficiency score, which is weighted with $1$ when an alpha of $1$ is chosen, leading to lower memory predictions.
In some cases, this can lead to very accurate predictions, while other tasks, such as those with a high variance in memory usage, may fail due to underprediction.
Switching between alpha parameters adaptively during workflow execution, as we do with the models, could address this problem and is an idea for future work. 

\section{Related Work}\label{sec:RELATED_WORK}
First, we discuss related methods focusing on workflow task memory prediction.
Second, we discuss related work that is similar to our proposed method but targets resource managers in general without a specific focus on workflows.

\subsection{Workflow Task Memory Prediction}

Substantial research has been done on predicting the memory requirements of workflow tasks.
We describe seminal work on this problem and compare each with our own method.

Da~Silva~et~al.~\cite{da2013toward} present a method that uses an online feedback loop to predict resource consumption, e.g., peak memory, of scientific workflow tasks.
Its prediction process uses a regression tree, built offline from historical monitoring data, that first classifies the workflow, application, and resource to be predicted.
Then, a data subset with high correlation is identified.
Their proposed method, Online-M, predicts the ratio between the I/O read and the resource metrics for correlated data and the median for uncorrelated data.
The extended method, Online-P~\cite{da2015online}, checks for a normal and gamma probability distribution in case there is no correlation identified it draws a value from these distributions.
Their online method includes updating task estimates upon completion of the workflow task.
Our method is also applied online and is able to update the models and thus the prediction results during the runtime of the workflow.
However, we use multiple models that are also able to capture non-linear memory consumption patterns and use a more selective failure handling and offsetting strategy to further reduce failed task executions caused by underestimation.

Witt~et~al.~\cite{witt2019feedback} propose a feedback-based resource allocation system for the scheduling of scientific workflows.
Their approach uses an online feedback loop to improve resource usage predictions and thus sizes resources for the tasks of a scientific workflow.
The authors examine two approaches to estimate the task's peak memory usage.
The first approach, the percentile predictor, predicts the task's peak memory usage by using certain percentiles, e.g., P50 for the median or P99 for the 99th percentile.
The second method is based on a linear regression model and estimates the task's peak memory consumption based on the relationship between input data size and memory usage.
The predictions of the linear model are then offset by the expected difference between the actual and the predicted peak memory usage.
Our work addresses a similar problem.
However, instead of relying on a single model class that is applied to all tasks, our method dynamically switches between model classes to find the most suitable for each task.

The peak memory allocation problem is also addressed in~\cite{witt2019learning}.
Here, again, the authors use the input data size to predict peak memory usage and train a prediction model that minimizes resource wastage instead of the prediction error.
They test quantile regression lines and select the parameters of the one with the least wastage.
Further, the authors examine different failure-handling strategies.
Again, we do not rely only on a single regression model class.
Sizey uses multiple regression models and includes dynamic switching on a per-task basis, allowing us to find the best model for heterogeneous tasks.

Tovar~et~al.~\cite{tovar2017job} present a job sizing strategy for tasks in scientific workflows that either minimizes wastage or maximizes throughput. 
The authors use the probabilities of peak resource values from the respective historical task traces to initialize a task.
Their goal is to minimize the sum of probabilities of resource peaks where the resource peak is greater than the allocated resource value.
The scope of our paper is similar since we also address the sizing of memory for tasks.
However, we also address task heterogeneity by including multiple parallel predictors while dynamically selecting the one with the best prediction results for each task.
Additionally, we propose a more selective retry strategy to avoid high resource wastage once a task fails.

Tovar~et~al.~\cite{tovar2022dynamic} propose another domain-specific approach that, instead of sizing the memory for tasks, aims to split the tasks into subtasks, matching a specific memory requirement.
Without historical task executions, the highest observed memory consumption is assigned to the first five tasks.
Upton task failure due to insufficient memory, the task is split into smaller tasks and resubmitted.
The initial size of a task is determined by a linear correlation between chunk size and task resource usage.
Contrary to our and the other presented approaches, the author's method does not directly predict the memory but rather the input size a task needs to consume a certain amount of memory.
The viability of such an approach depends on the structure and divisibility of the task input data and is thus not generally applicable to black-box tasks used in scientific workflows.

Bader~et~al.~\cite{bader2022leveraging} proposed two reinforcement learning methods based on gradient bandits and Q-learning.
The object of their reinforcement learning methods is the minimization between allocated and used memory while avoiding task failure.
The authors use no offsetting technique, as the reward functions implicitly discourage the agents from underestimating the resources.
One drawback of the proposed reinforcement learning methods is that they do not incorporate the dependency between task input size and resource usage, leading to higher wastage for tasks with fluctuating memory usage.

Bader~et~al.~\cite{baderPredictDynamicMemoryRequ2023} also presented a method to predict the memory consumption of workflow tasks over time.
To do this, they used time series monitoring data and predicted the expected task runtime.
The expected runtime is then divided into segments.
For each segment, they then trained a linear regression model that predicts peak memory, resulting in a step function that models a task's memory consumption over time.
Both the internal runtime prediction model and the segment-by-segment peak memory prediction models incorporate an offset strategy to avoid overall task failure.
In contrast to our method, which is a combination of multiple predictors, the resultant prediction function is a piecewise linear regression.
The method also requires time-series monitoring data, which may not be available in many workflow systems, and relies on the resource manager's ability to adjust memory allocations over time, a feature that is widely unavailable.

\subsection{Ensemble Memory Prediction for Resource Managers}

There is related work that has focused more generally on memory prediction for cluster workloads.
Similar to our workflow memory prediction method, closely related methods employ the best of multiple models.

Tanash~et~al.~\cite{tanash2019improving} proposed a Slurm extension that predicts a job's memory usage using machine learning models.
Their method intercepts the submission process by analyzing the job, predicting the expected memory usage, and adjusting the job limits.
While the prediction is done online, the actual model training is done offline and the model is not updated using completed jobs.
Tanash~et~al. decided to use the decision tree regression as it achieved the best prediction results.
Instead of selecting a single model to apply globally, Sizey dynamically selects the best performing model for each task.
In addition, Sizey works in a full online fashion, incorporating and updating the internal model at runtime to improve prediction accuracy by either updating the current model or even switching to a different model if it performs better with more data.

Rodrigues~et~al.\cite{rodrigues2016helping} present a memory prediction extension for the popular LSF HPC resource manager.
Their predictive method trains multiple machine learning models based on online and offline data collection and uses a sliding window to focus on the most recent submitted and finished jobs.
The machine learning models included are Support Vector Machines, Random Forests, Multilayer Perceptrons, and k-Nearest Neighbors.
Instead of addressing it as a regression problem, they discretize the memory predictions and turn it into a classification problem with memory sizes of 512MiB.
We approach this as a regression problem, which allows for more fine-grained predictions that can result in less wastage.
Further, our method allows switching between a weighting mechanism as presented by Rodrigues~et~al. and an Argmax strategy where the predictor with the highest resource allocation quality score is used for the prediction.
Our results show that such an Argmax strategy is beneficial in scenarios where the predictors are divergent, and weighting would bias the result.
Instead of relying solely on the accuracy of the validation data, we introduced an efficiency score that measures the deviations of the models from each other.

\section{Conclusion}\label{sec:CONCLUSION}
This paper presented Sizey, a novel online memory prediction method for workflow tasks that employs multiple machine learning models. 
The method dynamically selects the best-fitting model for each task based on an evaluation of each model's accuracy and memory prediction efficiency. 
To this end, Sizey incorporates monitoring information during the workflow execution and updates its prediction models while continuously re-evaluating the models' performance based on gathered data.

We implemented a prototype of Sizey and compared it to four state-of-the-art methods from the literature, using metrics from six real-world nf-core Nextflow workflows.
The results demonstrate that Sizey is able to significantly outperform the baselines, with a reduction in memory wastage of 24.68\% compared to the best state-of-the-art baseline.
Moreover, our experiments offer insights into the parameter selection of our method, thereby facilitating its deployment in other systems.

In the future, we plan to integrate our method into state-of-the-art online scheduling by considering the predicted required memory already during scheduling.

\section*{Acknowledgments}
{\small Funded by the Deutsche Forschungsgemeinschaft (DFG, German Research Foundation) as FONDA (Project 414984028, SFB 1404).}

\bibliographystyle{IEEEtran}
\balance
\bibliography{references}

\end{document}